\documentclass[10pt]{article}   
\pdfoutput=1
\usepackage{jheppub}
\usepackage{amsmath,amssymb,amsfonts,mathbbol,graphicx,slashed,color,amsthm, mathtools, upgreek, enumerate, tensor}

\usepackage{subcaption}
\usepackage{setspace}
\usepackage[export]{adjustbox}
\usepackage{arydshln}
\usepackage[dvipsnames]{xcolor}
\usepackage{physics}
\usepackage{hyperref}
\usepackage{comment}
\graphicspath{{pics/}}
\usepackage{breqn}
\usepackage{comment}
\usepackage{subcaption}

\usepackage{cleveref}

\usepackage{tikz}
\usetikzlibrary{shapes.misc}
\tikzset{cross/.style={cross out, draw=black, minimum size=2*(#1-\pgflinewidth), inner sep=0pt, outer sep=0pt}, cross/.default={1pt}}
\usetikzlibrary{decorations.pathmorphing}
\usetikzlibrary{arrows}

\allowdisplaybreaks

\colorlet{darkblue}{blue!70!black}

\colorlet{darkgreen}{green!50!black}
\colorlet{darkred}{red!50!black}

\def\bea{\begin{eqnarray}}
\def\eea{\end{eqnarray}}
\def\be{\begin{equation}}
\def\ee{\end{equation}}

\newcommand\GB[1]{{\it \color{darkgreen}  [#1 - GB]}}

\title{Timelike boundaries in de Sitter JT gravity and the Gao-Wald theorem}

\author[a]{Gauri Batra}
\affiliation[a]{Stanford Institute for Theoretical Physics, 382 Via Pueblo, Stanford, CA 94305}

\emailAdd{gb377@stanford.edu}
 
 \vspace{5mm}

\vspace{1cm}

\abstract{
The Gao-Wald theorem says that de Sitter spacetime reacts to a positive energy perturbation by getting ``taller." How does this change in the presence of timelike boundaries? We study this question in two-dimensional de Sitter JT gravity coupled to conformal matter. The effect of the boundaries has its roots in quantum corrections to the vacuum energy of the CFT due to Casimir-like effects. We consider two different kinds of timelike boundaries, either at locations where the coordinate $\varphi$ is constant or at locations where the dilaton is constant. For each kind we compute the vacuum expectation value of the matter CFT stress tensor. The stress tensor violates the null energy condition in the first case and saturates it in the second case, with each case exhibiting a negative energy density in the vacuum state. We then compute the semiclassical backreaction of this energy density and show how it can make the spacetime ``fatter" or ``taller" or a combination of both, depending on the regime. The spacetime getting fatter corresponds to an increase in the value of the dilaton at the horizon and hence also in the static patch horizon area in the higher-dimensional solution. We comment on the implications of our results for the quantum theories living on the timelike boundaries.
 }

\begin{document}

\maketitle
\parskip=10pt

\section{Introduction}\label{sec:intro}
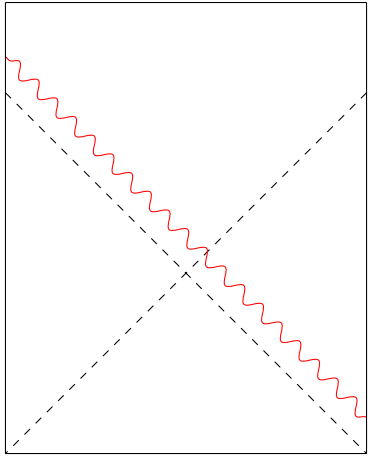
\begin{figure}
\centering
\begin{tikzpicture}[scale=0.6]
\node (I)    at ( 4,0) {};  % {I};

\node (II)   at (-4,0) {}; %  {II};

\path  
  (II) +(90:6)  coordinate  (IItop)
        +(90:4)  coordinate  (IImid)
       +(-90:4) coordinate (IIbot)
       +(90:4.8) coordinate (IIstart)
       +(0:4)   coordinate                  (IIright);
\draw[dashed]     (IImid) -- (IIright) -- (IIbot) ;
\path 
   (I) +(90:6)  coordinate (Itop)
   +(90:4)  coordinate  (Imid)
       +(-90:4) coordinate (Ibot)
       +(-90:3.2)  coordinate  (Iend)
       +(180:4) coordinate (Ileft);
\draw[dashed]  (Imid) -- (Ileft) -- (Ibot); 
\draw[red,decoration={snake},decorate]  (Iend) -- (IIstart);
% Zig zag singularity lines
\draw (IItop) to (Itop);
\draw (IItop) -- (IIbot);
\draw (Itop) -- (Ibot);   
\draw (IIbot) to (Ibot)  ;
\end{tikzpicture}
%\end{align}
\caption{In the presence of matter, de Sitter spacetime gets ``taller." As a result, a light ray (in red) sent sufficiently early from one pole reaches the other pole in finite time.}
\label{fig:taller}
\end{figure}

One intriguing feature of de Sitter spacetime is the effect of adding matter that satisfies the null energy condition, which according to the Gao-Wald theorem makes the spacetime ``taller" \cite{Gao_2000}\footnote{More precisely, the Gao-Wald theorem says that if a spacetime satisfies the null energy condition, the null generic condition, global hyperbolicity and null geodesic completeness, and moreover has a compact Cauchy surface, then the light cone of a point at sufficiently late but finite times will intersect an entire Cauchy surface in the past. Adding matter to dS leads to the null generic condition being satisfied.}. If we imagine de Sitter space as a sphere expanding with time, this theorem is also saying that in the presence of matter, signals sent sufficiently early from one pole will be able to reach the other pole in a finite amount of time (see Figure \ref{fig:taller}). This is in contrast to Anti de-Sitter space, where the addition of matter makes the spacetime ``fatter." Previous works that have explored this theorem in de Sitter spacetime include \cite{Anninos_2022, bousso2002adventuressitterspace, Leblond:2002ns, Leblond_2003, Levine_2023, Geng:2019bnn, Geng:2020kxh}.

This feature should have an avatar in any proposed holographic dual of de Sitter. Recent years have seen a large amount of progress in static patch holography, where the dual description lives either on a timelike boundary, the stretched horizon, or on the worldline of an observer \cite{Coleman:2021nor,susskind2021entanglement,Rahman:2022jsf,narovlansky2023doublescaled}. When we consider the full bulk theory with matter, we impose certain boundary conditions for the matter content on these surfaces. For quantum matter, this gives rise to a non-zero vacuum energy (for instance, due to moving mirror radiation or the Casimir effect). In case this vacuum energy violates the null energy condition, we can ask if the opposite of the Gao-Wald effect is true: if we consider the ground state of a quantum field theory on de Sitter space with timelike boundaries, do we see a fatter spacetime? Can we quantify this in some way? How is this reflected in the ground state and the first few excited states of the dual quantum mechanical theory? 

In this work we study the first two questions in a simple toy model. We consider two-dimensional Jackiw-Teitelboim (JT) gravity with positive cosmological constant coupled to conformal matter. This theory is obtained by dimensionally reducing the Nariai black hole spacetime \cite{maldacena2020dimensional, Svesko_2022}. The reduction gives rise to a dilaton, which measures the size of the transverse sphere in the higher-dimensional solution. The metric is that of dS$_2$:
\begin{align}
    ds^2=\frac{-d\sigma^2+d\varphi^2}{\cos^2\sigma},
\end{align}
which we will describe in more detail in Section \ref{sec:setup}.
We place timelike boundaries in this spacetime, on which the value of the dilaton is fixed. The effect of quantum matter in this setup is a non-zero stress tensor expectation value $\ev{T_{\mu\nu}}$ in the vacuum state. By studying the resulting semiclassical backreaction of this stress tensor, we answer the questions posed above. Specifically, we
find that the spacetime may get taller or fatter depending on the regime, even in cases where the stress tensor violates the null energy condition. There are also regimes where the spacetime gets taller for $\sigma$ in some interval, and fatter outside that interval. The regime in which the spacetime gets fatter corresponds to the vacuum energy causing an increase in the value of the dilaton at the horizon, and hence also in the static patch horizon area in the higher-dimensional solution. Here we single out the relation to the horizon area in the higher-dimensional solution since the dS horizon reacts in an interesting way to the addition of NEC-conforming matter to the static patch, shrinking in response to it \cite{Bousso_2000,Bousso:2000md,Maeda_1998,Chandrasekaran_2023}.

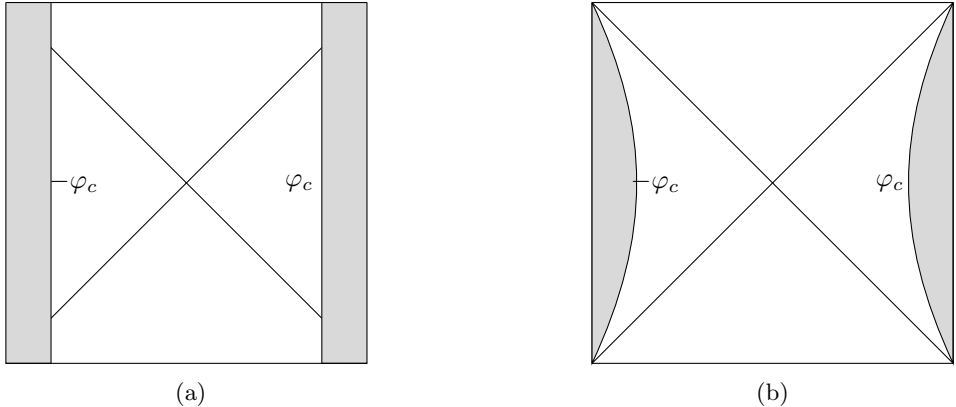
\begin{figure}
 \begin{subfigure}{0.49\textwidth}
 \centering
\begin{tikzpicture}[scale=0.6]
\node (I)    at ( 4,0) {};  % {I};
\node (II)   at (-4,0) {}; %  {II};
\path  
  (II) +(90:4)  coordinate  (IItop)
       +(-90:4) coordinate (IIbot)
       +(0:4)   coordinate                  (IIright);
\draw     (IItop) -- (IIright) -- (IIbot) ;
\path 
   (I) +(90:4)  coordinate (Itop)
       +(-90:4) coordinate (Ibot)
       +(180:4) coordinate (Ileft);
\draw  (Itop) -- (Ileft) -- (Ibot); 
\draw[fill=gray!30]  (3,4) -- (3,-4) -- (Ibot) -- (Itop) ;
\draw[fill=gray!30]  (-3,4) -- (-3,-4) -- (IIbot) -- (IItop) ;
% Zig zag singularity lines
\draw (IItop) to (Itop); 
\draw (IIbot) to (Ibot)  ;
\node at (2.5,0) {$\varphi_c$};
\node at (-2.5,0) {$-\varphi_c$};
\end{tikzpicture}
 \label{fig:boundary1}
 \caption{}
\end{subfigure}
 \begin{subfigure}{0.49\textwidth}
 \label{fig:boundary2}
 \centering
\begin{tikzpicture}[scale=0.6]
\node (I)    at ( 4,0) {};  % {I};
\node (II)   at (-4,0) {}; %  {II};
\path  
  (II) +(90:4)  coordinate  (IItop)
       +(-90:4) coordinate (IIbot)
       +(0:4)   coordinate                  (IIright);
\draw     (IItop) -- (IIright) -- (IIbot) ;
\path 
   (I) +(90:4)  coordinate (Itop)
       +(-90:4) coordinate (Ibot)
       +(180:4) coordinate (Ileft);
\draw  (Itop) -- (Ileft) -- (Ibot); 
% Zig zag singularity lines
\draw (IItop) to (Itop);
\draw (IItop) -- (IIbot);
\draw[fill=gray!30] (Itop) to[bend right=25] (Ibot) -- (Itop);
\draw[fill=gray!30] (IItop) to[bend left=25] (IIbot) -- (IItop);
\draw (Itop) -- (Ibot);
\draw (IIbot) to (Ibot)  ;
\node at (2.6,0) {$\varphi_c$};
\node at (-2.6,0) {$-\varphi_c$};
\end{tikzpicture}
 \caption{}
\end{subfigure}
%\end{align}
\caption{The two kinds of boundaries we will consider in this work: (1) at constant $\varphi=\varphi_c$ (left) and (2) at curves of constant dilaton given by $\cos\varphi_c=\tanh\zeta\cos\sigma$ (right). The grey region corresponds to parts of the spacetime that have been cut out.}
\label{fig:boundaries_penrose}
\end{figure}

Timelike boundaries in de Sitter space have been considered and studied previously in numerous works \cite{Coleman:2021nor, Batra:2024kjl, Svesko_2022, Banihashemi:2022htw, Anninos:2024wpy, Geng:2021wcq}. The timelike boundaries we consider in this work are of two types based on the dilaton profile at the boundaries. Boundaries of type I are placed at constant $\varphi=\pm\varphi_c$, on which the dilaton profile is fixed to be $\sigma$-dependent. Boundaries of type II are placed at the curves $\cos\varphi_c=\tanh\zeta \cos\sigma$, on which the dilaton is fixed to a constant. Both types of boundaries are restricted to lie in the range $\varphi\in(-\frac{\pi}{2},\frac{\pi}{2}]$. They are pictured in Figure \ref{fig:boundaries_penrose}. The second type of boundaries are more relevant if we are interested in boundaries placed at constant radial position $r$ in the static patch in higher dimensions, such as in the works \cite{Coleman:2021nor, Batra:2024kjl}. The motivation for considering the first type is that the computation of the Casimir energy $\ev{T_{\mu\nu}}$ is simpler and serves as a good warmup for the second case.

The outline of this paper is as follows. In Section \ref{sec:setup} we describe the dS$_2$ JT gravity setup in more detail, along with descriptions of the two types of timelike boundaries. In Section \ref{sec:criterion} we explain why we can look at the Gao-Wald theorem through the lens of backreaction, and lay out the criterion for the spacetime getting taller or fatter. 
In Sections \ref{sec:bdry1} and \ref{sec:bdry2} we compute the expectation value of the stress energy tensor of the matter CFT in its vacuum state for the two different kinds of boundaries. We find the resulting backreaction on the spacetime and answer the defining question of whether the spacetime gets taller or fatter, and by how much. We conclude in Section \ref{sec:discussion} with a summary and some future directions. The Appendices include details of calculations that we do not include in the main text. We set $l_{\mathrm{dS}}=1$ in the rest of the paper.

\section{Setup}\label{sec:setup}
In this paper we work in $2$D de Sitter JT gravity coupled to a matter CFT, obtained by dimensionally reducing the Nariai black hole spacetime \cite{maldacena2020dimensional,Svesko_2022}. The relevant terms in the action are given by
\begin{align}
    S=\frac{1}{4\pi}\left[\int_{\Sigma_2} \phi(R-2)+2\int_{\partial\Sigma_2^t}\phi K\right]+S_{\mathrm{CFT}},
\end{align}
where we added boundary terms for the timelike boundaries. $\phi$ is the dilaton, which represents the area of the transverse sphere in the higher-dimensional theory. Varying with respect to $\phi$ gives $R=2$ and fixes the metric to
\begin{align}
\label{eq:metric}
    ds^2=\frac{-d\sigma^2+d\varphi^2}{\cos^2\sigma},
\end{align}
where $\sigma$ ranges from $-\frac{\pi}{2}$ to $\frac{\pi}{2}$ and $\varphi$ from $-\pi$ to $\pi$. We will mainly be interested in the range $\varphi\in(-\frac{\pi}{2},\frac{\pi}{2}]$, for which the dilaton goes to $+\infty$ at $\sigma=\pm\frac{\pi}{2}$. The horizon is located at the curve $\cos\varphi=\cos\sigma$, and the poles are located at $\varphi=\pm\frac{\pi}{2}$.
Varying the above action with respect to $g_{\mu\nu}$ gives
\begin{align}
\label{eq:covariant_eom}
    (g_{\mu\nu}\nabla^2-\nabla_\mu\nabla_\nu+g_{\mu\nu})\phi=2\pi T_{\mu\nu}.
\end{align}
Setting $T_{\mu\nu}=0$ gives the vacuum solution for the dilaton as
\begin{align}
    \phi=\phi_r\frac{\cos\varphi}{\cos\sigma},
\end{align}
where $\phi_r$ represents the area of the horizon in the higher-dimensional theory.
In this work, we will consider the semiclassical backreaction of quantum matter, and upgrade $T_{\mu\nu}$ to $\ev{T_{\mu\nu}}$.
We can then write Eq. (\ref{eq:covariant_eom}) as the following three coupled differential equations:
\begin{align}
\label{eq:dilaton_eom}
    &\left(-\partial_\varphi^2+\tan\sigma\partial_\sigma-\sec^2\sigma\right)\phi=2\pi \ev{T_{\sigma\sigma}},\\
    &\left(-\partial_\sigma^2+\tan\sigma\partial_\sigma+\sec^2\sigma\right)\phi=2 \pi \ev{T_{\varphi\varphi}},\\
    &\left(-\partial_\sigma\partial_\varphi+\tan\sigma\partial_\varphi\right)\phi=2 \pi \ev{T_{\sigma\varphi}}.
\end{align}
Using conservation of the matter stress energy tensor $\nabla_\mu \ev{T^{\mu\nu}}=0$, we can find the general solution to these equations that is symmetric under $\sigma\to -\sigma$ and $\varphi\to-\varphi$ (derived in Appendix \ref{sec:deoms}):
\begin{align}
\label{eq:dilaton_soln}
    \phi=\phi_r'\frac{\cos\varphi}{\cos\sigma}-2\pi\sec\sigma\int^\sigma d\sigma' \cos\sigma' \int^{\sigma'}d\sigma'' \ev{T_{\varphi\varphi}}\nonumber,
\end{align}
where $\phi_r'$ is an arbitrary integration constant. It may be fixed by specifying additional data, such as initial conditions for the dilaton. Considering this two-dimensional setup is helpful because the metric (\ref{eq:metric}) is related to 2D Minkowski space by a conformal factor, which makes the computation of $\ev{T_{\mu\nu}}$ of the matter CFT in its ground state particularly simple. We will focus on the case where $c/\phi_r<<1$, so that matter backreaction is small.

\begin{comment}
\begin{figure}
\centering
\begin{tikzpicture}[scale=0.8]
\node (I)    at ( 4,0) {};  % {I};
\node (II)   at (-4,0) {}; %  {II};
\path  
  (II) +(90:4)  coordinate  (IItop)
       +(-90:4) coordinate (IIbot)
       +(0:4)   coordinate                  (IIright);
\draw     (IItop) -- (IIright) -- (IIbot) ;
\path 
   (I) +(90:4)  coordinate (Itop)
       +(-90:4) coordinate (Ibot)
       +(180:4) coordinate (Ileft);
\draw  (Itop) -- (Ileft) -- (Ibot); 
% Zig zag singularity lines
\draw (IItop) to (Itop);
\draw (IItop) -- (IIbot);
\draw (Itop) -- (Ibot);   
\draw (IIbot) to (Ibot)  ;
\end{tikzpicture}
\label{fig:penrose}
\caption{}
\end{figure}
\end{comment}

\subsection{Brown-York stress tensor and matter boundary conditions}\label{sec:by}
Before computing the bulk stress tensor and its backreaction, we study the gravitational system more by noting the expressions for the Brown-York stress tensor for the boundaries. For this purpose, we will model the matter CFT in this subsection as \cite{Almheiri:2014cka,Svesko_2022} 
\begin{align}
    S_{\mathrm{CFT}}=-\frac{c}{24\pi}\int_{\Sigma_2}(\partial_\mu\chi\partial^\mu\chi+\chi R)-\frac{c}{12\pi}\int_{\partial\Sigma_2^t}  \chi K,
\end{align}
where the last term is a boundary term for the timelike boundaries.
The terms relevant for finding the Brown York Hamiltonian of the timelike boundaries may be written as
\begin{align}
\label{eq:grstuff}
    S_{\mathrm{BY}}&=\frac{1}{4\pi}\int_{\Sigma_2} \phi R+\frac{1}{2\pi}\int_{\partial\Sigma_2^t}\phi K-\frac{c}{24\pi}\int_{\Sigma_2}\chi R-\frac{c}{12\pi}\int_{\partial\Sigma_2^t}  \chi K\\
    &=\frac{1}{4\pi}\left[\int d\sigma d\varphi \sqrt{-g} \partial^\mu\left(\phi-\frac{c}{6}\chi\right) (g_{\mu\rho}\partial_\sigma-g_{\sigma\rho}\partial_\mu)g^{\sigma\rho}\right].
\end{align}
In going from the first to the second line we used the analysis in \cite{Almheiri:2014cka}. Let $w$ be the coordinate transverse to the boundary and $\tau$ be the coordinate parallel to it. The Brown York Hamiltonian is then given by the Hamilton Jacobi result, which relates the variation of the action with respect to the boundary metric to the conjugate momentum associated with $g^{\tau\tau}$:
\begin{align}
\label{eq:hj}
    T_{\tau\tau}^{\mathrm{BY}}=-\frac{2}{\sqrt{-h}}\frac{\delta S}{\delta h^{\tau\tau}}=\frac{2}{\sqrt{-h}}\frac{\partial L}{\partial_w(g^{\tau\tau})}.
\end{align}
In this work we study two different types of timelike boundaries. In the first case we place them at locations where $\varphi$ is constant, at $\varphi=\pm\varphi_c$. In this case, the value of the dilaton at the boundaries is $\sigma$-dependent. To find the Brown York stress tensor, it is easiest to work in the coordinates in Eq. (\ref{eq:metric}), with $\sigma$ being the coordinate along the boundary and $\varphi$ being the coordinate transverse to the boundary. Using Eq. (\ref{eq:hj}), the Brown York Hamiltonian for one boundary at $\varphi=\varphi_c$ is then
\begin{align}
    T_{\sigma\sigma}^{\mathrm{BY}}=\frac{1}{2\pi}\sec\sigma\partial_\varphi\left(\phi-\frac{c}{6}\chi\right)|_{\varphi=\varphi_c}.
\end{align}
Note that in general the above tensor does not satisfy $\nabla_\sigma T^{\mathrm{BY},\sigma\sigma}=0$. One can derive special boundary conditions for $\chi$ for which this is true. However, such boundary conditions won't be invariant under conformal transformations. Since we use this invariance to compute $\ev{T_{\mu\nu}}$, we won't consider these boundary conditions in this work.

The second type of timelike boundaries are placed at locations where the dilaton $\phi$ is constant on the boundaries. To find the Brown York Hamiltonian in this case, it is convenient to transform from the metric (\ref{eq:metric}) to the coordinates
\begin{align}
    ds^2=d\rho^2-\cos^2\rho dt^2,
\end{align}
where $\sin\rho=\frac{\cos\varphi}{\cos\sigma}$. These coordinates cover one static patch, with the horizon at $\rho=\frac{\pi}{2}$ and the pole at $\rho=0$. The boundary is located at $\rho=\rho_c$, so that $\rho$ is the coordinate transverse to the boundary and $t$ is the coordinate along the boundary. The Brown York Hamiltonian for one boundary at $\rho=\rho_c$ is then
\begin{align}
    T_{tt}^{\mathrm{BY}}=\frac{1}{2\pi}\cos^2\rho_c\partial_\rho\left(\phi-\frac{c}{6}\chi\right)\Big|_{\rho=\rho_c}.
\end{align}
As mentioned earlier, in this work we consider conformal boundary conditions for the matter content, which include $\chi|_{\rho_c}=0$ and $\partial_\rho \chi|_{\rho_c}=0$. These boundary conditions are saying that the flux of matter out of the system is zero, so that the gravitational system is "boxed in". We will also impose that $\partial_\rho \chi|_{\rho_c}$ is a constant, which ensures the covariant conservation of the Brown York Hamiltonian computed above, $\nabla_t T^{\mathrm{BY},tt}=0$.

\section{Criterion for tallness or fatness}\label{sec:criterion}
\subsection{Probing the Gao-Wald theorem via backreaction}\label{sec:gaowald}
In this work we study the Gao-Wald theorem in the presence of timelike boundaries in a simple toy model by finding the backreaction of quantum matter. It is useful to outline first why we can look at spacetime getting taller or fatter through this lens of backreaction. Tallness can be understood either as the Penrose diagram of de Sitter space getting taller on adding matter, or as two causally disconnected parts of de Sitter space getting causally connected on adding matter (both are evident in Figure \ref{fig:taller}).

Let us first concretely relate the dS Penrose diagram getting taller to the backreaction of matter on the geometry. Restricting to $2+1$ dimensions, we can parameterize the change in the global de Sitter metric due to the addition of spherically symmetric matter as (setting $l_{\mathrm{dS}}=1$)
\begin{align}
    ds^2=-H_t(t,\varphi)^2dt^2+H_\varphi(t,\varphi)^2\cosh^2t(d\varphi^2+\sin^2\theta d\theta^2).
\end{align}
Note that $t\in(-\infty,\infty)$. The functions $H_t$ and $H_\varphi$ encode the matter backreaction along with various boundary or initial conditions. To find the Penrose diagram at constant $\theta$, we can define $\Omega=H_\varphi\cosh t$ and $
H_t dt=\Omega d\sigma$. This transforms the relevant part of the metric to
\begin{align}
    ds^2=\Omega^2(-d\sigma^2+d\varphi^2).
\end{align}
Moreover, integrating both sides of $H_t dt=\Omega d\sigma$ gives
\begin{align}
\label{eq:tsigma}
    \int dt\frac{ H_t(t,\varphi)}{H_\varphi(t,\varphi)}\sech t=\int d\sigma.
\end{align}
To find the range of $\sigma$, we do the $t$ integral from $-\infty$ to $\infty$.  For $H_\varphi=H_t=1$ corresponding to global de Sitter, we get a square Penrose diagram since both $\sigma$ and $\varphi$ take values in $(-\frac{\pi}{2},\frac{\pi}{2})$. If $\frac{H_t}{H_\varphi}>1$ for all $t$ and $\varphi$, we get a taller Penrose diagram since the range of $\sigma$ increases, and fatter otherwise. The value of $\frac{H_t}{H_\varphi}$ is sensitive to various properties of matter, along with the boundary and initial conditions.

\begin{figure}
 \begin{subfigure}{0.49\textwidth}
 \centering
\begin{tikzpicture}[scale=0.6]
\node (I)    at ( 4,0) {};  % {I};
\node (II)   at (-4,0) {}; %  {II};
\path  
  (II) +(90:4)  coordinate  (IItop)
       +(-90:4) coordinate (IIbot)
       +(0:4)   coordinate                  (IIright);
\draw[<-]     (IItop)+(-0.5,0.5) -- (IIright) -- (IIbot) ;
\path 
   (I) +(90:4)  coordinate (Itop)
       +(-90:4) coordinate (Ibot)
       +(180:4) coordinate (Ileft);
\draw[<-]  (Itop)+(0.5,0.5) -- (Ileft) -- (Ibot); 
\draw[red,decoration={snake},decorate]  (Ibot) -- (-4,2);
\node at (-4.8,4.8) {$U$};
\node at (4.8,4.8) {$V$};
% Zig zag singularity lines
\draw (IItop) to (Itop); 
\draw (IItop) to (IIbot);
\draw (Itop) to (Ibot);
\draw (IIbot) to (Ibot)  ;
\end{tikzpicture}
 \label{fig:boundary1a}
 \caption{}
\end{subfigure}
 \begin{subfigure}{0.49\textwidth}
 \label{fig:boundary2a}
 \centering
\begin{tikzpicture}[scale=0.6]
\node (I)    at ( 4,0) {};  % {I};
\node (II)   at (-4,0) {}; %  {II};
\path  
  (II) +(90:4)  coordinate  (IItop)
       +(-90:4) coordinate (IIbot)
       +(0:4)   coordinate                  (IIright);
\draw[<-]     (IItop)+(-0.5,0.5) -- (IIright) -- (IIbot);
\path 
   (I) +(90:4)  coordinate (Itop)
       +(-90:4) coordinate (Ibot)
       +(180:4) coordinate (Ileft);
\draw[<-]  (Itop)+(0.5,0.5) -- (Ileft) -- (Ibot); 
\draw[red,decoration={snake},decorate]  (Ibot) -- (-2.5,4);
\node at (-4.8,4.8) {$U$};
\node at (4.8,4.8) {$V$};
% Zig zag singularity lines
\draw (IItop) to (Itop);
\draw (IItop) -- (IIbot);
\draw (Itop) -- (Ibot);
\draw (IIbot) to (Ibot)  ;
\end{tikzpicture}
 \caption{}
\end{subfigure}
%\end{align}
\caption{A picture of understanding de Sitter space getting taller (fatter) in terms of the two poles becoming causally connected (disconnected). On the left is the taller spacetime, and on the right is the fatter spacetime. The wiggly red line represents a null geodesic sent from the right to the left pole. For the taller spacetime, this null particle speeds up, so that $V(U=\infty)<0$ and the two sides become causally connected. The opposite effect happens on the right, so that $V(U=\infty)>0$. Note this is not a conformal diagram, since the wiggly red lines are null geodesics. One would have to make the diagram taller on the left or fatter on the right for the wiggly red lines to travel at $45^o$.}
\label{fig:UV}
\end{figure}
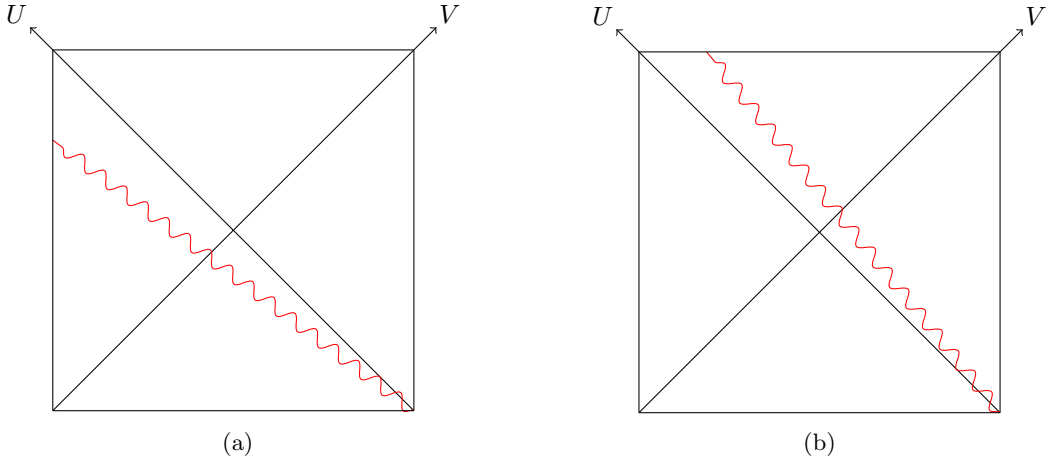

The tallness illustrated above leads to two causally disconnected parts of de Sitter space becoming causally connected, as pictured in Figure \ref{fig:taller}. We can concretely relate this phenomenon to the stress tensor satisfying the averaged null energy condition (ANEC) as follows. The argument is a small modification of the AdS argument outlined in \cite{Gao_2017}. We will consider three-dimensional de Sitter space in the following coordinates:
\begin{align}
    ds^2=\frac{1}{(1-UV)^2}(-4dUdV+(1+UV)^2 d\theta^2).
\end{align} 
If we add a matter source to this system, we can parameterize the backreaction by the perturbed metric
\begin{align}
    ds^2=\frac{1}{(1-UV)^2}(-4dUdV+h_{UU}dU^2+(1+h_{\theta\theta})(1+UV)^2 d\theta^2).
\end{align}
Consider a null particle geodesic that leaves the right pole at $U=-\infty$. This geodesic reaches the left pole at $V(U=\infty)$. To tell whether a perturbation makes the spacetime fatter or taller, we use the sign of $V(\infty)$, which is approximately given by
\begin{align}
    V(\infty)=-(g_{UV}(0))^{-1}\int_{-\infty}^{\infty}dUh_{UU}.
\end{align}
$V(\infty)>0$ indicates a fatter geometry, and corresponds to the particle slowing down. So, a fatter spacetime corresponds to
\begin{align}
    \int_{-\infty}^{\infty}dUh_{UU}>0.
\end{align}
For a taller spacetime we have
\begin{align}
    \int_{-\infty}^{\infty}dUh_{UU}<0,
\end{align}
indicating that the particle reaches the other side in a finite time. These two cases are illustrated in Figure \ref{fig:UV}.
The above integral involving $h_{UU}$ is related to the ANEC through one of Einstein's equations evaluated at $V=0$:
\begin{align}
    -\frac{1}{2}h_{UU}-\frac{1}{2} \partial_U(Uh_{UU})-\frac{1}{2} \partial_U^2h_{\theta\theta}=\ev{T_{UU}}.
\end{align}
If we assume that the $h_{UU}$ and $\partial_U h_{\theta\theta}$ go to zero as $U\to\pm\infty$ (which is valid if the perturbation is small), integrating from $U=-\infty$ to $U=\infty$ gives
\begin{align}
    \int_{-\infty}^{\infty}dU h_{UU}=-2\int_{-\infty}^{\infty}dU\ev{T_{UU}}.
\end{align}
We see that if the ANEC is satisfied (that is, if $\int_{-\infty}^{\infty}dU\ev{T_{UU}}>0$), then the LHS of the above equation is negative, leading to a taller spacetime. If the ANEC is violated, then the opposite is true.

\subsection{Criterion in JT gravity} \label{sec:jt_criterion}
In JT gravity, the backreaction of matter is encoded in the change in configuration of the dilaton, with the metric (\ref{eq:metric}) staying unaffected. This is in contrast to the higher-dimensional case, where the backreaction is encoded in the change in the metric. For this reason, we need a criterion for fatness or tallness which is different from that outlined in Section \ref{sec:gaowald}, but still maintains the spirit of tallness being related to causal connectability\footnote{I thank Edgar Shaghoulian for emphasizing this.}. Additionally, we are looking for a criterion that is covariant.

Before we describe the criterion, note that the locations of the timelike boundaries in general will be different in the backreacted solution compared to the unbackreacted solution. This is because the dilaton is held fixed on the boundaries. Since the dilaton configuration changes as a function of $\sigma$ and $\varphi$ due to a non-zero $\ev{T_{\mu\nu}}$, so do the locations of the boundaries, since they track locuses of fixed dilaton configuration.

We are then interested in the following measure to judge fatness or tallness: if the boundaries become more causally connected due to backreaction, we say the spacetime gets taller, and fatter otherwise. More precisely, consider sending a signal from the right boundary to the left boundary at some boundary time $\sigma_R$. The time $\sigma_L$ at which this signal arrives at the left boundary changes due to backreaction, since the boundary configurations change. If $\sigma_L$ decreases, so that the signal reaches the right boundary earlier, the spacetime gets taller, and fatter if $\sigma_L$ increases. Note that this criterion is in general $\sigma$-dependent. 

\begin{figure}
\centering
\begin{tikzpicture}[scale=0.6]
\node (I)    at ( 4,0) {};  % {I};
\node (II)   at (-4,0) {}; %  {II};
\path  
  (II) +(90:4)  coordinate  (IItop)
       +(-90:4) coordinate (IIbot)
       +(0:4)   coordinate                  (IIright);
\draw     (IItop) -- (IIright) -- (IIbot) ;
\path 
   (I) +(90:4)  coordinate (Itop)
       +(-90:4) coordinate (Ibot)
       +(180:4) coordinate (Ileft);
\draw  (Itop) -- (Ileft) -- (Ibot); 
% Zig zag singularity lines
\draw (IItop) to (Itop);
\draw (IItop) -- (IIbot);
\draw (Itop) to[bend right=18] (Ibot) ;
\draw (IItop) to[bend left=18] (IIbot) ;
\draw[color=magenta] (Itop) to[bend right=45] (Ibot) ;
\draw[color=magenta] (IItop) to[bend left=45] (IIbot) ;
\draw (Itop) -- (Ibot);
\draw (IIbot) to (Ibot)  ;
\draw[red,decoration={snake},decorate] (2.4,0) -- (-1.6,4);
\draw[red,decoration={snake},decorate] (3.3,0) -- (-0.7,4);
\node at (-1.6,4.5) {$\varphi_s'$};
\node at (-0.7,4.4) {$\varphi_s$};
\end{tikzpicture}
\caption{An example of the criterion outlined in Section \ref{sec:jt_criterion} for initially disconnected boundaries (black) that are also disconnected after backreaction (pink), for a signal sent from $\sigma_R=0$. $\varphi_s$ goes to $\varphi_s'$ on taking backreaction into account. In this case, we see that we obtain a taller spacetime since the proper distance between $\varphi_s$ and $\varphi=-\frac{\pi}{2}$ decreases on backreaction.}
\label{fig:const_criterion}
\end{figure}
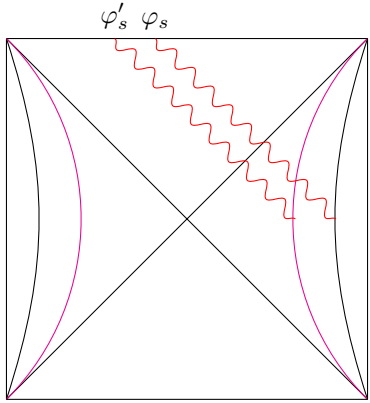

Applying this criterion is trickier in the case where the boundaries are initially causally disconnected (such as the second kind of boundaries in Figure \ref{fig:boundaries_penrose}b). In the vacuum solution, any signal sent from the left boundary does not reach the right boundary. Due to backreaction, one possibility is that this signal reaches the left boundary at $\sigma_L<\frac{\pi}{2}$, clearly giving a taller solution. In case backreaction does not lead to this signal reaching the left boundary, we can no longer compare $\sigma_L$ between the backreacted and unbackreacted solutions. We need a different condition to distinguish tallness and fatness, which is as follows. Let $\varphi_s$ be the location where the signal from the right boundary intersects $\sigma=\frac{\pi}{2}$. Consider the proper distance between $\varphi_s$ and $\varphi=-\frac{\pi}{2}$.  If this distance decreases due to backreaction, we obtain a taller solution, and fatter otherwise. This is visualised in Figure \ref{fig:const_criterion}.

In the next few sections we will compute $\ev{T_{\mu\nu}}$ explicitly for various boundary configurations and use this criterion in practice.

\begin{comment}
\GB{Another possible criterion is the following: consider two curves at locations of constant dilaton. If we send out a signal at a time $\sigma$ from one of these curves to the second curve, does it arrive faster or slower in the backreacted solution? The first corresponds to tallness, the second to fatness. Note these curves would be at a different locus in $(\sigma,\varphi)$ in the backreacted solution, which is why this time would change.}
\end{comment}

\section{Boundaries at constant $\varphi$}\label{sec:bdry1}
In this section we consider a matter CFT living on a dS$_2$ strip with timelike boundaries at $\varphi=\pm\varphi_c$ (Figure \ref{fig:boundaries_penrose}a), with conformal boundary conditions on the matter. These are boundary conditions that are invariant under conformal transformations, and include Neumann ($\partial_\varphi\chi|_{\pm\varphi_c}=0$) and Dirichlet ($\chi|_{\pm\varphi_c}=0$) boundary conditions. The metric on the strip is given by Eq. (\ref{eq:metric}), which is the flat metric $\hat g$ multiplied by a Weyl factor $e^{2\omega}=\sec^2\sigma$. We are interested in computing $\ev{T_{\mu\nu}}$ for this matter CFT and understanding whether it makes the geometry taller or fatter. This is inspired by a similar computation carried out in the absence of boundaries in \cite{Levine_2023}.

\subsection{Computation of $\ev{T_{\mu\nu}}$} \label{sec:tds_comp}

We can compute $\ev{T_{\mu\nu}}$ by using the transformation properties of the stress tensor under a Weyl rescaling in the presence of boundaries \cite{Herzog_2016}:
\begin{align}
\label{eq:weyl}
    \ev{T_{\mu\nu}}_{g}=\ev{T_{\mu\nu}}_{\hat{g}}-\frac{c}{12\pi}\left(\hat{\nabla}_\mu\omega\hat{\nabla}_\nu\omega-\frac{1}{2}\hat{g}_{\mu\nu}(\hat{\nabla}\omega)^2-\hat{\nabla}_\nu\hat{\nabla}_\mu\omega+\hat{g}_{\mu\nu}\hat{\nabla}^2\omega\right)+\frac{c}{12\pi}\delta(x^{\perp})h_{\mu\nu}n^\rho\partial_\rho\omega.
\end{align}
In the above formula, $\delta(x^\perp)$ is the Dirac delta function with support on the boundaries, $h_{\mu\nu}$ is the induced metric on the boundary of interest, and $n^\rho$ is the normal vector to it. $\ev{T_{\mu\nu}}_{\hat{g}}$ is the stress tensor of the CFT on a strip of length $2\varphi_c$ with a flat metric. We can calculate it via conformal transformation from the upper half plane (UHP) \cite{cardy2008boundary}, as described below. We will work in complex coordinates $z$ and $\bar{z}$.

\begin{figure}
    \centering
    \begin{tikzpicture}
        \draw[color=darkgreen,line width=5pt]  (-3,0) -- (0,0) ;
        \draw[color=orange,line width=5pt]  (0,0) -- (3,0);
        \node at (3.5,0) {$\Re(z)$};
            \draw (0,0) -- (0,3) node [below right] {$\Im(z)$};
            \fill[gray!30]  (-3,0) -- (3,0)  --  (3,-1.5) -- (-3,-1.5);
        \draw[->] (3.5,1.3) -- (5,1.3) node at (4.2,1) {$w=\frac{2\varphi_c}{\pi}\log z$};
        \draw[color=orange,line width=5pt]  (5.2,0) -- (10.8,0) ;
        \draw[color=darkgreen,line width=5pt]  (5.2,2) -- (10.8,2) ;
            \fill[gray!30] (5.2,0) -- (10.8,0) --  (10.8,-1.5) -- (5.2,-1.5);
            \fill[gray!30] (5.2,2) -- (10.8,2) --  (10.8,3) -- (5.2,3);
        \node at (11.3,0) {$\Re(w)$};
        \node at (11.8,2) {$\Im(w)=2\varphi_c$};
    \end{tikzpicture}
    \caption{The conformal map from the upper half plane to the strip of width $2\varphi_c$. The shaded grey regions are removed, and lines that are mapped to each other are outlined in the same color.}
    \label{fig:map}
\end{figure}
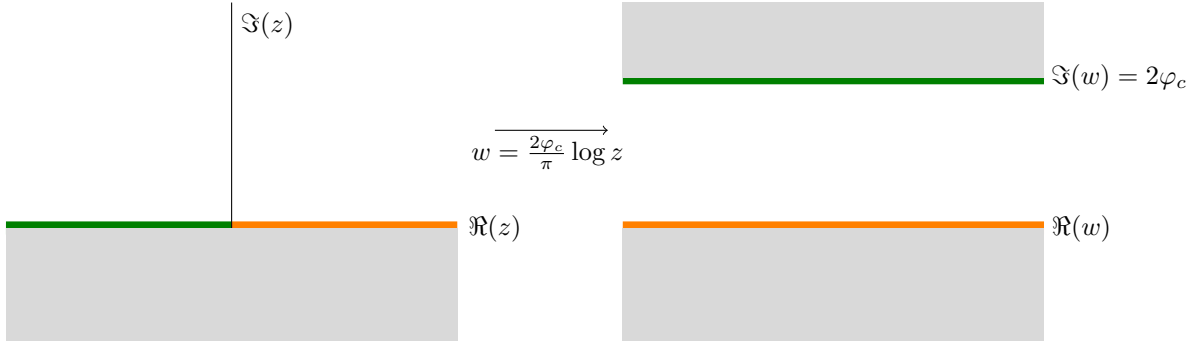

The UHP is the restriction of the complex plane to $\mathrm{Im}(z)>0$, so that the real axis $\mathrm{Im}(z)=0$ is the boundary (see left panel of Figure \ref{fig:map}). We consider boundary conditions such that $\ev{T}=\ev{\bar{T}}$ or $T_{\sigma\varphi}=0$ at $\mathrm{Im}(z)=0$, which translates to the energy-momentum flux out of the strip being zero. These include the conformal boundary conditions discussed above. To find $\ev{T_{\mu\nu}}$ on the strip, we need the stress tensor one-point function on the UHP, which equals zero. One might have expected an expression for $\ev{T_{\mu\nu}}$ that depends on the distance from the boundary, since the boundary breaks the translation symmetry in the direction transverse to it. However, it is not possible to construct a tensor that respects the remaining unbroken symmetries (which include scale invariance and translations parallel to the boundary).

To find the stress tensor on the strip from that on the UHP, we need the map from the UHP to the strip. Equipping the strip with the complex $w$ coordinate, this map is given by
\begin{align}
    w=\frac{2\varphi_c}{\pi}\log{z},
\end{align}
as visualized in Figure \ref{fig:map}.
The stress tensor one-point function on the strip may then be found using the transformation of the stress tensor under a conformal map, and doing a coordinate transformation back to the $(\sigma,\varphi)$ coordinates:
\begin{align}
    \ev{T_{\mu\nu}}_{\hat g}=-\frac{c\pi}{96\varphi_c^2}\delta_{\mu\nu}.\label{eq:strip}
\end{align}
Using Eq. (\ref{eq:weyl}), we get the full stress tensor on the dS$_2$ strip as
\begin{align}
\label{eq:tmunu1}
    \ev{T_{\mu\nu}}_{g}=\left(\frac{c}{24\pi}-\frac{c\pi}{96\varphi_c^2}\right)\delta_{\mu\nu}+\frac{c}{24\pi}g_{\mu\nu}.
\end{align}
Note that when the boundaries are at the poles at $\varphi_c=\pm\frac{\pi}{2}$, we get an answer that is proportional to $g_{\mu\nu}$. As the boundaries move away from the poles, the stress tensor gets an extra negative contribution due to the Casimir effect,
\begin{align}
\label{eq:tildeT1}
    \ev{T_{\mu\nu}}_{g}-\frac{c}{24\pi}g_{\mu\nu}=\left(\frac{c}{24\pi}-\frac{c\pi}{96\varphi_c^2}\right)\delta_{\mu\nu}<0.
\end{align}
A covariant measure of this negativity is the fact that the energy density measured by an inertial observer is negative:
\begin{align}
    \ev{T_{\mu\nu}}_{g} u^{\mu} u^{\nu}<0,
\end{align}
where $u^\mu$ is a timelike vector. The stress tensor expectation value we computed also violates the null energy condition.

\subsection{Backreaction of $\ev{T_{\mu\nu}}$ on the dilaton and boundaries}
As discussed earlier, the backreaction of the non-zero $\ev{T_{\mu\nu}}$ computed above is encoded in the change in the dilaton configuration as well as the timelike boundary location. In this subsection we will explicitly compute this change in dilaton configuration. This is then tied to the boundaries moving to different locations $\pm\varphi_c^n$ in the backreacted solution, which we will also compute later in this subsection. 

We can solve for the new value of the dilaton $\phi_n$ by using Eq. (\ref{eq:dilaton_soln}) and the expression for $\ev{T_{\mu\nu}}$ computed in the last subsection:
\begin{align}
\label{eq:newphi1}
    \phi_n=\phi_r'\frac{\cos\varphi}{\cos\sigma}+\frac{c\pi^2}{48\varphi_c^2}\left(1+\sigma\tan\sigma\right)-\frac{c}{12}\sigma\tan\sigma,
\end{align}
where in general $\phi_r'$ may be different from $\phi_r$. It may be fixed by (for instance) initial conditions for the dilaton. Recall that the horizon is given by $\cos\varphi=\cos\sigma$, so $\phi_r$ represents the area of the horizon without any matter in the higher-dimensional solution.
We can look at this solution in two regimes:
\begin{enumerate}
    \item $\phi_r'-\phi_r>-\frac{c\pi^2}{48\varphi_c^2}$: For these values of $\phi_r'$, the dilaton value increases at the horizon at $\sigma=0$ (and also other values of $\sigma$). 
    \item $\phi_r'-\phi_r<-\frac{c\pi^2}{48\varphi_c^2}$: For these values of $\phi_r'$, the dilaton value decreases at the horizon at $\sigma=0$.
\end{enumerate}
Since we are considering the backreaction of a stress tensor that violates the null energy condition (NEC), and we expect NEC-conforming matter to make the higher-dimensional dS horizon smaller \cite{Bousso_2000,Bousso:2000md,Maeda_1998,Chandrasekaran_2023}, we might expect the first regime to be more physical. It would be interesting to understand more whether such considerations from higher dimensions may be used to constrain the values that $\phi_r'$ can take. 

The above change in the dilaton also implies that the timelike boundaries are located at different values of $\varphi$ in the backreacted solution. This is because we impose the boundary condition that the value of the dilaton on the timelike boundaries is fixed to its vacuum value. Their new locations $\varphi=\pm\varphi_c^n$ are then determined by equating the value of the dilaton on the boundaries in the backreacted solution with its value on the boundaries in the vacuum solution:
\begin{align}
    &\phi_n(\varphi_c^n)=\phi(\varphi_c)\\
    \implies & \phi_r'\frac{\cos\varphi_c^n}{\cos\sigma}+\frac{c\pi^2}{48{\varphi_c}^2}\left(1+\sigma\tan\sigma\right)-\frac{c}{12}\sigma\tan\sigma=\phi_r\frac{\cos\varphi_c}{\cos\sigma}. \label{eq:newloc}
\end{align}
Note $\phi_r'-\phi_r$ is proportional to $c$ in the above equation, since it should equal zero in the absence of matter. Solving this equation gives a $\sigma$-dependent answer for $\varphi_c^n$.

\begin{figure}
    \centering
    \includegraphics[width=2.8in]{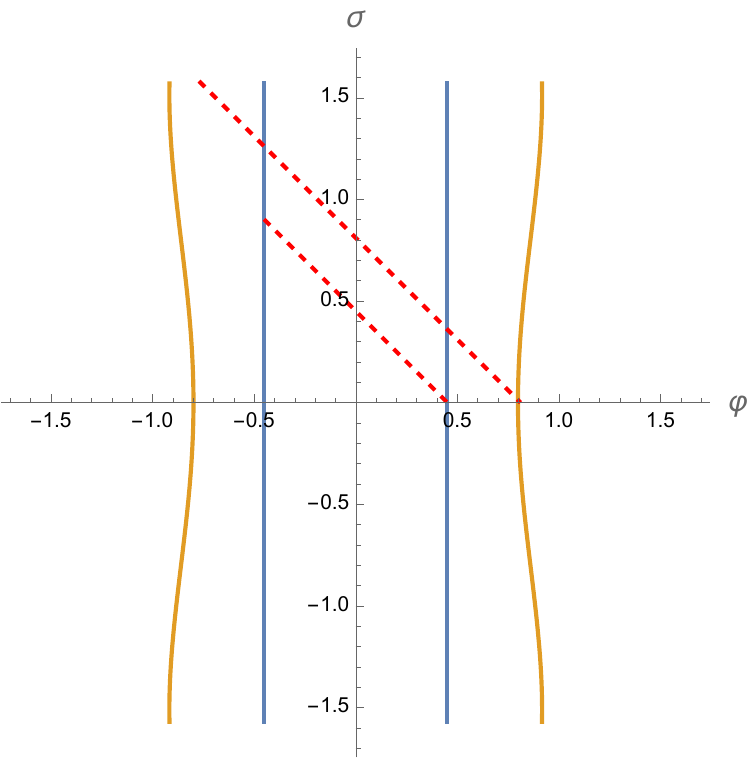}
    \caption{A picture of the locations of the timelike boundaries before (blue) and after (orange) backreaction, for $\phi_r'=\phi_r=5$ and $c=1$. The boundaries in the unbackreacted solution are at constant $\varphi_c=\pm\cos^{-1}(0.9)$. Backreaction leads to a fatter solution since a signal (red) sent from the left boundary either takes a longer time to reach the right boundary or does not reach it.}
    \label{fig:boundaries_after_1}
\end{figure}

Let us focus on the regime with $\phi_r'-\phi_r>-\frac{c\pi^2}{48\varphi_c^2}$, for which the value of the dilaton increases at the horizon. Since $\cos x$ is a decreasing function of $x$ in the range we are looking at, from the above equation we see that $\varphi_c^n>\varphi_c$ for the two sides to be equal. This translates to an increase in the proper distance between the boundaries. Additionally, we get a fatter geometry according to the criterion of Section \ref{sec:jt_criterion}, since a signal sent from the right boundary reaches the left boundary at a later time. For $\phi_r'=\phi_r$ and $c=1$, these effects are visualised in Figure \ref{fig:boundaries_after_1}. 

\begin{figure}
    \centering
    \includegraphics[width=2.8in]{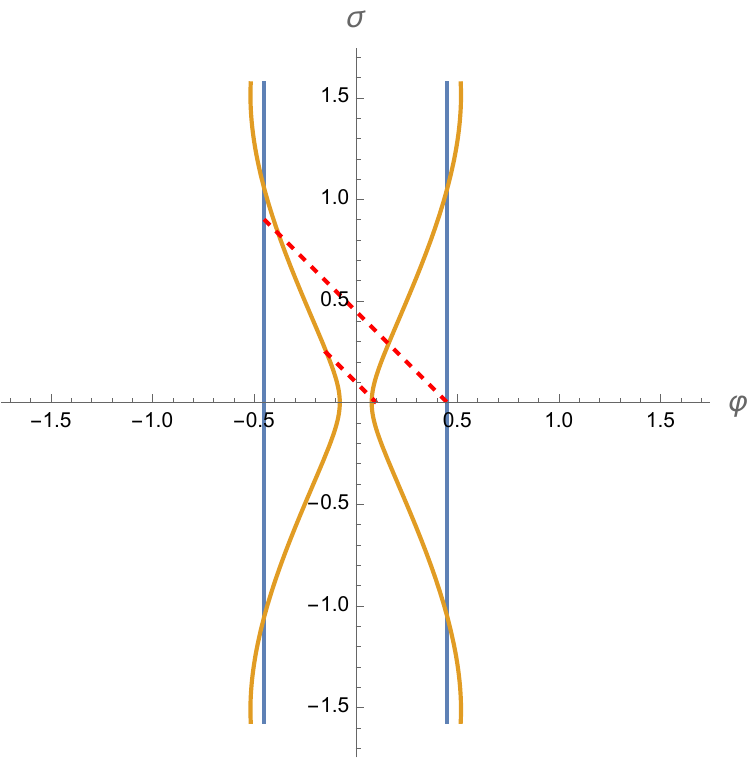}
    \includegraphics[width=2.8in]{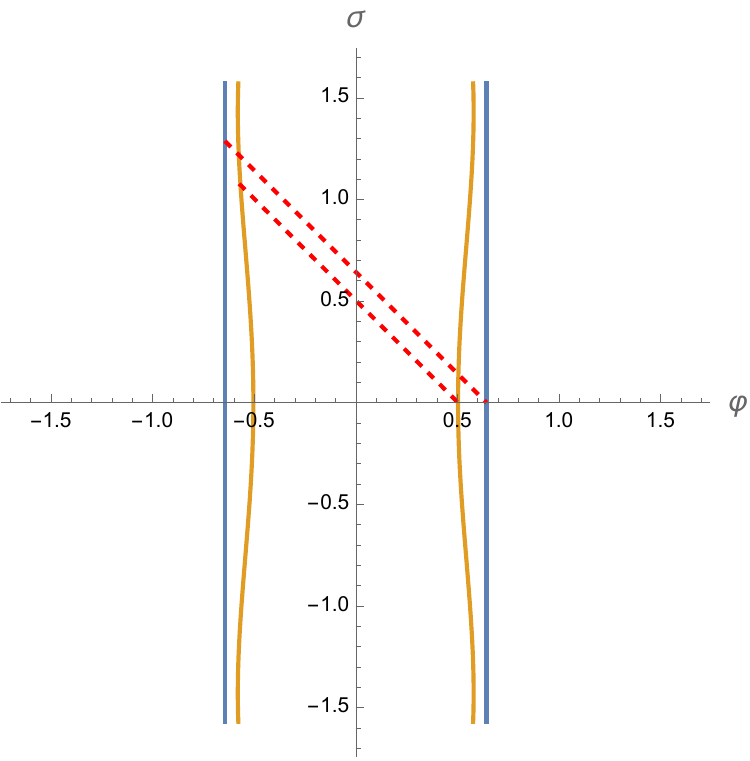}
    \caption{A picture of the locations of the timelike boundaries before (blue) and after (orange) backreaction, for $\phi_r=5$ and $c=1$. $\phi_r'=3.5$ for the left figure and $\phi_r'=4$ for the right figure. The boundaries in the unbackreacted solution are at constant $\varphi_c=\pm\cos^{-1}(0.9)$ in the left figure and $\varphi_c=\pm\cos^{-1}(0.8)$ in the right figure. On the left we see a taller solution in some interval in $\sigma$, and on the right we see a taller solution overall. In red is a signal from the left boundary that reaches the right boundary faster due to backreaction, indicating tallness.}
    \label{fig:taller_boundaries_1}
\end{figure}

For the second regime with $\phi_r'-\phi_r<-\frac{c\pi^2}{48\varphi_c^2}$, more complex phenomena are possible. The geometry may get taller for some interval in $\sigma$ and fatter outside that interval, or it may get taller overall. Both these examples are pictured in Figure \ref{fig:taller_boundaries_1}. Note that there also exist values of $\phi_r'$ for which there is no consistent timelike embedding of the boundaries.

\subsection{Higher order effects}
In the above analysis we found $\ev{T_{\mu\nu}}$ for a given boundary configuration, and then found that its backreaction causes the proper distance between the boundaries to change. One subtlety is that when this happens, the value of $\ev{T_{\mu\nu}}$ changes again, causing the proper distance between the boundaries to change again. Below we will argue that this effect is subleading for $c/\phi_r<<1$, the regime we are working in.

First note from Eq. (\ref{eq:newloc}) that for $c/\phi_r<<1$, the new location of the boundaries is shifted by an amount that is $\mathcal{O}(c/\phi_r)$. This means that the stress tensor for the new boundary locations $\ev{T_{\mu\nu}}(\varphi_c^n)$  is different from $\ev{T_{\mu\nu}}(\varphi_c)$ by an amount that is $\mathcal{O}(c^2/\phi_r)$, since $\ev{T_{\mu\nu}}$ is proportional to $c$.
This would correspond to adding a correction to Eq. (\ref{eq:newloc}) that is $\mathcal{O}(c^2/\phi_r)$. The resulting shift in the location of the boundaries is then $\mathcal{O}(c/\phi_r)^2$, which is subleading. 

To get the exact result, we would need to solve the equation
\begin{align}
    \phi_r'\frac{\cos\varphi_c^n}{\cos\sigma}+f(T_{\mu\nu}(\varphi_c^n),\sigma,\varphi)=\phi_r\frac{\cos\varphi_c}{\cos\sigma}.
\end{align}
Note the second term on the left hand side is also dependent on $\varphi_c^n$, as opposed to Eq. (\ref{eq:newloc}). $f(T_{\mu\nu}(\varphi_c^n),\sigma,\varphi)$ is determined from the general solution to the dilaton equations of motion given in Eq. (\ref{eq:dilaton_soln}). This equation is in general hard to solve because we don't have a result for $T_{\mu\nu}(\varphi_c^n)$ when the boundary locations are arbitrary, which is why we restrict to the regime where $c/\phi_r<<1$.

\section{Boundaries at constant dilaton}\label{sec:bdry2}
In this section we will consider a matter CFT living on dS$_2$ with metric given by Eq. (\ref{eq:metric}), and timelike boundaries at locations given by the curves
\begin{align}
    \cos\varphi_c=\tanh\zeta \cos\sigma.
\end{align}
These boundaries are pictured on the Penrose diagram in Figure \ref{fig:boundaries_penrose}b. They are more relevant from the perspective of higher-dimensional constructions of holographic duals to de Sitter \cite{Coleman:2021nor,Batra:2024kjl}, where the boundary is at a fixed radial location inside the static patch. Note that the boundaries move closer together on increasing $\zeta$. They are causally disconnected, one in each static patch. We are again interested in computing $\ev{T_{\mu\nu}}$ for this matter CFT, and checking whether it makes the geometry taller or fatter. As explained in Section \ref{sec:by}, the boundary conditions on the matter content satisfy two conditions: (1) they are invariant under conformal transformations, and (2) they maintain conservation of the Brown York Hamiltonian.

\subsection{Computation of $\ev{T_{\mu\nu}}$} \label{sec:radial_comp}
To compute $\ev{T_{\mu\nu}}$ in this case, we cannot pursue the strategy of Section \ref{sec:bdry1}, since there is no obvious conformal transformation from the upper half plane that leads to boundaries at the above locations. To use CFT tools for computing $\ev{T_{\mu\nu}}$, we will first map the global $(\sigma,\varphi)$ coordinates to the $(\sigma',\varphi')$ coordinates. These coordinates satisfy two conditions. The first is that the boundaries in these coordinates are located at constant $\varphi=\pm\frac{\pi}{2}$, so that we may use the expression (\ref{eq:strip}) for $\ev{T_{\mu\nu}}$ of a matter CFT on a flat strip. The second is that the metric is related to flat space by a Weyl factor, so we may use Eq. (\ref{eq:weyl}) to find the stress tensor on this metric from the flat space answer.

The $(\sigma',\varphi')$ coordinates are given by\footnote{The way we found this transformation was by realizing that boosts in AdS$_3$ act in a similar way on such curves on the asymptotic boundary, taking $\frac{\cos\phi}{\cos t}=\tanh\zeta$ to constant $\phi'$. The explicit boost transformations in AdS$_3$ are given in \cite{Lindgren:2015fum}. We also thank Gonzalo Torroba for noting that this is similar to the Casini-Huerta-Myers transformation in \cite{Casini_2011}.}
\begin{align}
\label{eq:transformation}
    \tan\sigma'&=\frac{\sin\sigma}{\cos\sigma\cosh\zeta-\cos\varphi\sinh\zeta},\\
    \cot\varphi'&=\frac{-\cos\sigma\sinh\zeta+\cos\varphi\cosh\zeta}{\sin\varphi}.
\end{align}
One can check that in these coordinates the boundary curves $\cos\varphi_c=\tanh\zeta \cos\sigma$ go to $\cot\varphi_c'=0$ (or $\varphi_c'=\pm\frac{\pi}{2}$). 
The metric (\ref{eq:metric}) is transformed to
\begin{align}
    ds^2_{g'}=e^{2\Omega(\sigma',\varphi')}(-d{\sigma'}^2+d{\varphi'}^2),
\end{align}
where the Weyl factor $e^{2\Omega(\sigma',\varphi')}$ is given implicitly in terms of $\sigma$ and $\varphi$ as
\begin{align}
    e^{2\Omega(\sigma',\varphi')}=\sec ^2\sigma \left(\sinh ^2\zeta  \cos (\sigma -\varphi ) \cos (\sigma +\varphi )-\sinh (2 \zeta ) \cos \sigma  \cos \varphi +\cosh ^2\zeta \right).
\end{align}
Using Eq. (\ref{eq:weyl}) for the Weyl transformation of the stress tensor, we may find $\ev{T_{\mu\nu}}_{g'}$ on this metric:
\begin{align}
\label{eq:static}
    \ev{T_{\mu\nu}}_{g'}=\ev{T_{\mu\nu}}_{\hat{g}}-\frac{c}{12\pi}\left(\hat{\nabla}_\mu\Omega\hat{\nabla}_\nu\Omega-\frac{1}{2}\hat{g}_{\mu\nu}(\hat{\nabla}\Omega)^2-\hat{\nabla}_\nu\hat{\nabla}_\mu\Omega+\hat{g}_{\mu\nu}\hat{\nabla}^2\Omega\right)+\frac{c}{12\pi}\delta(x^{\perp})h_{\mu\nu}n^\rho\partial_\rho\Omega,
\end{align}
where the last term has support only on the boundaries, and would be unimportant for our analysis.
Since this transformation also maps the two boundaries to $\varphi_c'=\pm\frac{\pi}{2}$, $\ev{T_{\mu\nu}}_{\hat{g}}$ is the stress tensor on the flat strip of width $\pi$, as computed in Eq. (\ref{eq:strip}):
\begin{align}
    \ev{T_{\mu\nu}}_{\hat{g}}=-\frac{c}{24\pi}\delta_{\mu\nu}.
\end{align}
We can then find $\ev{T_{\mu\nu}}_{g}$ from $\ev{T_{\mu\nu}}_{g'}$ by coordinate transforming back to $(\sigma,\varphi)$ coordinates. This calculation is performed numerically in Mathematica. We start by outlining the numerical results in the two static patches for $c=1$ and $\zeta=1$. In the static coordinates, the form of $\ev{T_{\mu\nu}}_{g}$ is particularly simple. For this we transform the expression for $\ev{T_{\mu\nu}}_{g}$ to the $(t,\rho)$ coordinates introduced in Section \ref{sec:by}:
\begin{align}
    ds^2=d\rho^2-\cos^2\rho dt^2,
\end{align}
which cover one static patch. $\rho=\frac{\pi}{2}$ corresponds to the horizon, and $\rho=0$ to the pole. $\zeta=1$ corresponds to the boundary being located at $\rho_c\approx 0.87$.
Since these coordinates are static and the boundary conditions on the matter field are imposed at time-independent locations, we find a time-independent answer for $\ev{T_{tt}}_g$ and $\ev{T_{\rho\rho}}_g$. We also find $\ev{T_{t\rho}}_g\approx 0$. At $\rho=0.9$, the curves for $\ev{T_{tt}}_g$ (in blue), $\ev{T_{\rho\rho}}_g$ (in orange) and $\ev{T_{t\rho}}_g$ (in green) look like the following as a function of time:
\begin{equation}
        \includegraphics[width=2.93in]{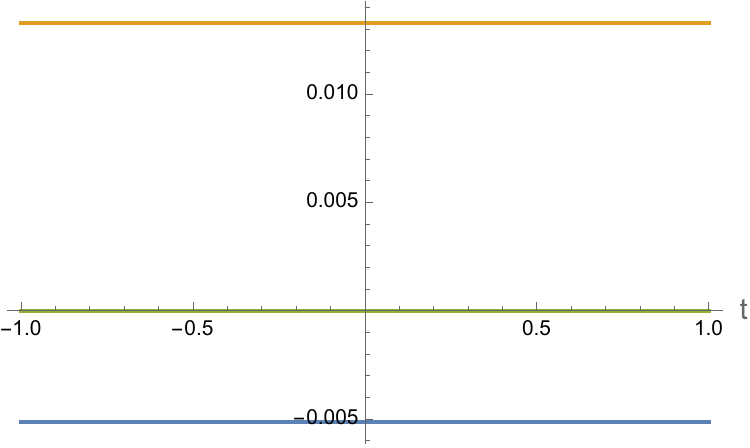}.
\end{equation}
As a function of $\rho$, $\ev{T_{\rho\rho}}_g$ (in orange below) is approximately constant at $\frac{c}{24\pi}$,
and $\ev{T_{tt}}_g$ (in blue) approximately equals $-\cos^2\rho \ev{T_{\rho\rho}}_g$:
\begin{equation}
        \includegraphics[width=2.93in]{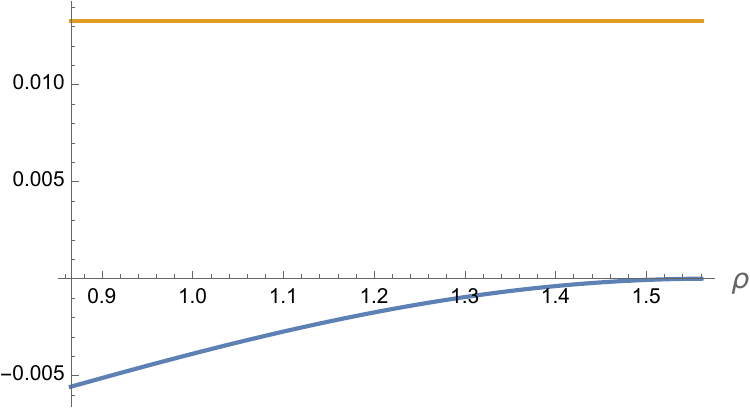}.
\end{equation}
This is also implied by $\nabla_\mu \ev{T^{\mu\nu}}_g=0$.
Note that the resulting energy density measured by any inertial observer in the two static patches is negative:
\begin{align}
   \ev{T_{\mu\nu}}_{g} u^{\mu} u^{\nu}<0,
\end{align}
where $u^\mu$ is any timelike vector. 
However, an interesting property of this $\ev{T_{\mu\nu}}_{g}$ is that it approximately saturates the null energy condition (NEC). 

In Section \ref{sec:bdry1} we found that the Casimir effect due to the boundaries at constant $\varphi$ violates the NEC. One possible reason why the NEC is not violated in this setup is that in this case, the boundaries are $\sigma$-dependent and accelerating (see Appendix \ref{sec:acceleration}). Placing boundary conditions at time-dependent locations leads to particle creation, which manifests itself in the form of positive ``moving mirror" radiation \cite{Davies:1976hi}. This positive radiation then competes with the negative energy density due to the Casimir effect, giving an overall saturation of the NEC. The reason the NEC is saturated is that the radiation propagates at the speed of light. 

Another surprising feature is that the bulk part of $\ev{T_{\mu\nu}}_g$ we computed is approximately independent of $\zeta$ inside the static patches, so that $\ev{T_{\rho\rho}}_g\approx\frac{c}{24\pi}$ (its value in the case where the boundaries are at $\varphi_c=\pm\frac{\pi}{2}$). It would be interesting to understand whether there is a symmetry at the heart of this behavior, and if the matter being conformal plays a role.

The stress tensor expectation value starts depending strongly on $\zeta$ at some $\zeta$-dependent point outside the static patches. The form of $\ev{T_{\mu\nu}}_g$ and its dependence on $\zeta$ outside the two static patches won't be required for the computation of the change in boundary location. For completeness, we detail its features in Appendix \ref{sec:tmunu}. These features indicate that the total energy of the matter CFT, found by integrating the energy density on a spatial slice, becomes less negative on increasing $\zeta$. This is another signal of the ``moving mirror" radiation.

\begin{comment}
We can check this by looking at the sign of $\tilde T_{\sigma\varphi}=T_{\sigma\varphi}$ for positive and negative $\varphi$ in Figure, which indicates that there is flux in the negative $\varphi$ direction from the right boundary and in the positive $\varphi$ direction from the left boundary. 
\end{comment}

\begin{comment}
One can see this by computing the magnitude of the acceleration four-vector $a_{\mu}a^{\mu}$. For this we parameterize the boundary path as $\sigma_c(\tau)$ and $\varphi_c(\tau)$, whose forms are fixed by the following two equations:
\begin{align}
    &\cos\varphi_c(\tau)=\tanh\zeta\cos\sigma_c(\tau),\\
    &-\left(\frac{d\sigma_c}{d\tau}\right)^2+\left(\frac{d\varphi_c}{d\tau}\right)^2=-\cos^2\sigma.
\end{align}
The second equation is the condition that the trajectory is timelike. The acceleration four vector is then given by 
\begin{align}
    a^{\mu}=\left(\frac{d^2\sigma_c}{d\tau^2},\frac{d^2\varphi_c}{d\tau^2}\right),
\end{align}
and has a non-zero magnitude squared $g_{\mu\nu} a^\mu a^\nu$. 
\end{comment}

\subsection{Backreaction of $\ev{T_{\mu\nu}}$ on the dilaton and boundaries}
Similar to Section \ref{sec:bdry1}, we want to analyse the change in the configurations of the dilaton and the boundaries due to the backreaction of $\ev{T_{\mu\nu}}_g$ computed above. To find the effect on the dilaton, it is again easier to work in the $(t,\rho)$ coordinates describing each static patch. We will use the approximate result for $\ev{T_{\mu\nu}}_g$ computed in the previous subsection:
\begin{align}
    \ev{T_{\rho\rho}}_g\approx \frac{c}{24\pi},\ \ \ev{T_{tt}}_g\approx -\frac{c}{24\pi}\cos^2\rho, \ \ \ev{T_{t\rho}}_g\approx 0.
\end{align}
Since $\ev{T_{\mu\nu}}_g$ is independent of $t$ and $\ev{T_{t\rho}}_g\approx 0$, the change in the dilaton configuration is also independent of $t$. The relevant equations of motion for the dilaton in this time-independent case become, using Eq. (\ref{eq:dilaton_eom}):
\begin{align}
    &(\cos^2\rho\partial_\rho^2+\cos^2\rho)\phi=-2\pi \ev{T_{tt}},\\
    &(\tan\rho\partial_\rho-1)\phi=-2\pi \ev{T_{\rho\rho}}.
\end{align}
Using $\nabla_\mu T^{\mu\nu}=0$, one can show that the solution to one equation also solves the other equation. For $T_{\mu\nu}=0$, the solution for the dilaton is $\phi=\phi_r\sin\rho$. The solution for $\phi$ for general $\ev{T_{\mu\nu}}$ may be written as
\begin{align}
    \phi=\phi_r'\sin\rho-2\pi\sin\rho\int^\rho d\rho'\cot\rho' \mathrm{cosec}\rho' \ev{T_{\rho\rho}},
\end{align}
where again in general $\phi_r'\neq \phi_r$. $\phi_r'$ may also depend on $\zeta$.  For the case of interest where $\ev{T_{\rho\rho}}=\frac{c}{24\pi}$, we have the dilaton solution
\begin{align}
    \phi_n=\phi_r'\sin\rho+\frac{c}{12}.
\end{align}
Similar to the analysis below Eq. (\ref{eq:newphi1}), we can analyze this solution in two regimes:
\begin{enumerate}
    \item $\phi_r'-\phi_r>-\frac{c}{12}$: For these values of $\phi_r'$, the dilaton value increases at the horizon.
     \item $\phi_r'-\phi_r<-\frac{c}{12}$: For these values of $\phi_r'$, the dilaton value decreases at the horizon.
\end{enumerate}
We are considering the backreaction of matter that approximately saturates the null energy condition. So, it is not clear which regime we might expect to be more physical, based on the effects of NEC-conforming matter on the area of the horizon in the higher-dimensional solution. Note that $\phi_r'$ may be fixed either by specifying initial conditions for the dilaton, or other constraints on the system. For instance, one could require that the change in the Brown York energy equals the energy of the matter content in the bulk.

\begin{figure}
    \centering
    \includegraphics[width=2.5in]{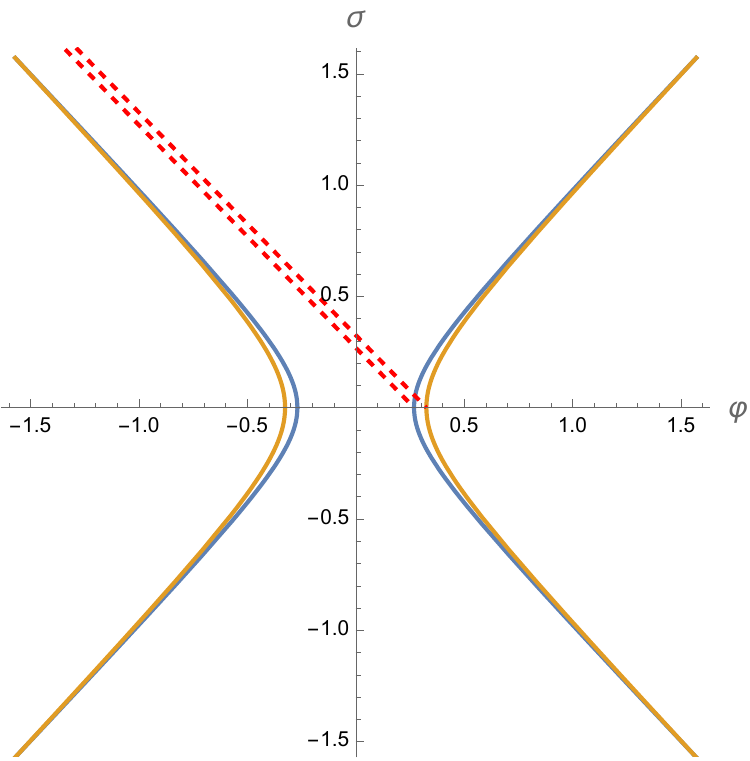}
    \includegraphics[width=2.5in]{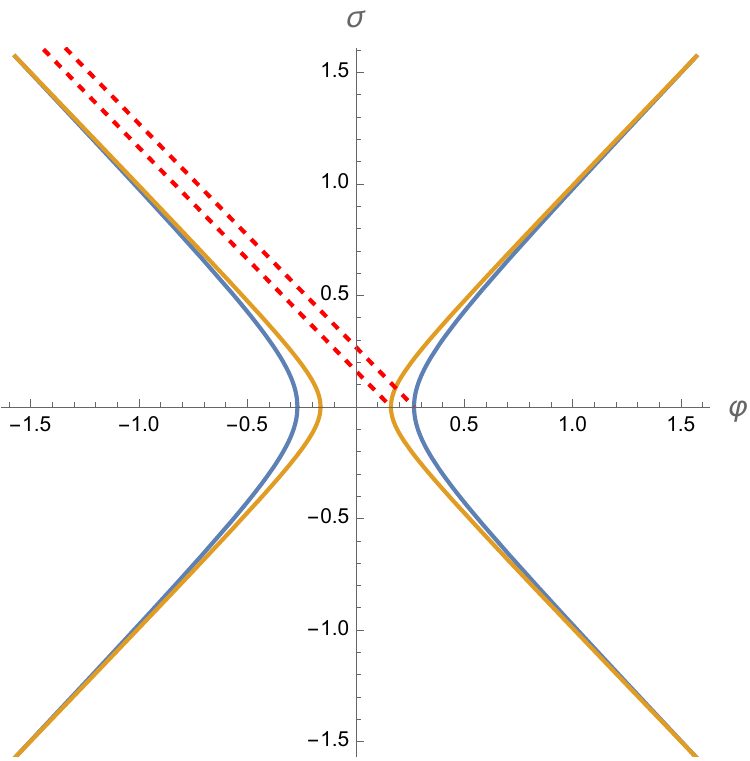}
    \caption{A picture of the locations of the timelike boundaries before (blue) and after (orange) backreaction, for $\phi_r=5$ and $c=1$. $\phi_r'=\phi_r=5$ for the left figure and $\phi_r'=4.8$ for the right figure. The boundaries in the unbackreacted solution are at $\varphi_c=\tanh\zeta \cos\sigma$, with $\zeta=2$. The signals sent from one boundary to the other are indicated in red, showing the solution on the left (right) gets fatter (taller) according to the criterion in Section \ref{sec:jt_criterion}.}
    \label{fig:boundaries_after_2}
\end{figure}

Using this backreacted solution for the dilaton, we can find the new location of the timelike boundary in one static patch, given by $\rho=\rho_c^n$. Since we fix the value of the dilaton on the boundary, its new location is given by equating the values of the dilaton on the boundary with and without taking backreaction into account. This translates into the condition
\begin{align}
    &\phi_n(\rho_c^n)=\phi(\rho_c)\\
    \implies & \phi_r'\sin\rho_c^n+\frac{c}{12}=\phi_r\sin\rho_c.
\end{align}
For the case where $\phi_r'-\phi_r>-\frac{c}{12}$, this means that $\rho_c^n<\rho_c$, implying that the boundary moves further away from the horizon. The effect on the location of the timelike boundary in the complementary static patch is the same due to symmetry, leading to an increase in the proper distance between the boundaries. Additionally, the criterion outlined in Section \ref{sec:jt_criterion} points to a fatter spacetime, since the proper distance to the left boundary of a signal sent from the right boundary is larger after backreaction. These effects are visualised in the left panel of Figure \ref{fig:boundaries_after_2} for $\phi_r'=\phi_r$ and $c=1$. The same reasoning implies that in the second regime where $\phi_r'-\phi_r<-\frac{c}{12}$, we may get a taller geometry, as pictured in the right panel of Figure \ref{fig:boundaries_after_2}. In this case too, there exist values of $\phi_r'$ for which there is no consistent timelike embedding of the boundaries.

\section{Discussion and implications for the boundary theory}\label{sec:discussion}
In this work we considered two-dimensional de Sitter JT gravity coupled to conformal matter. We studied the effect of timelike boundaries placed symmetrically in this spacetime on the causal properties of the vacuum state. In particular, we computed the vacuum expectation value of the stress tensor of the CFT in the presence of two kinds of boundaries. The first type is located at constant $\varphi=\pm\varphi_c$, and the second at locuses where the dilaton is constant, so that $\cos\varphi_c=\tanh\zeta\cos\sigma$. In both cases, we found a negative energy density in the vacuum state of the CFT. In the first case, the stress tensor violates the null energy condition, and approximately saturates it in the second case (as indicated by numerics).

Our defining question was whether the opposite of the Gao-Wald effect occurs in the presence of the negative and possibly NEC-violating vacuum energy due to the boundaries: does the spacetime get fatter instead?
We showed that either effect is possible, and that the spacetime can get taller or fatter depending on the value of a parameter $\phi_r'$, which is an integration constant in the solution for the dilaton. It may be fixed using additional physical input or initial conditions for the dilaton, and is related to the value of the dilaton at the horizon. If $\phi_r'$ is such that the dilaton increases at the horizon due to the vacuum energy, we obtain a fatter solution. If $\phi_r'$ is such that the dilaton decreases at the horizon at $\sigma=0$, it is also possible to get a taller solution, or a solution that is taller in some interval in $\sigma$ and fatter outside that interval. Here, tallness is defined as the two boundaries becoming more causally connected, and fatness as the two boundaries becoming more causally disconnected.

The future directions opened up by our work are outlined below.
\begin{enumerate}
    \item Similar effects are expected if the timelike boundaries are replaced by observers, since we impose boundary conditions at the observer locations. The key difference is that we do not cut out the spacetime that exists beyond the observers (unlike for timelike boundaries, as in Figure \ref{fig:boundaries_penrose}). One question we could ask is how our results for whether the spacetime gets taller or fatter change on taking this into account. What happens if we consider multiple observers?
    
    \item It would be interesting to consider the implications of our results for the quantum theories that live on the timelike boundaries or observer worldlines. Without taking the quantum effects of matter into account, one might have expected that adding a tiny perturbation would make the spacetime taller (based on the Gao Wald theorem), leading to the quantum theories on the two boundaries being coupled to one another. 

    In the cases where quantum effects make the spacetime fatter in the vacuum state of the matter, we would encounter a qualitative difference when coupling the two theories. To counteract the effect of the negative energy, we would have to add a finite perturbation instead of an infinitesimal perturbation to couple them. One question is if our bulk computation of the Casimir energy provides a prediction for a lower bound on how big this perturbation has to be.
    On the other hand, in the cases where the quantum effects makes the spacetime taller in the vacuum state of the matter, one has to consider the coupling of the theories on the two boundaries in their ground states. 

    \item The causal connectability (or lack thereof) of the two boundaries should be encoded in the singularities of two-sided correlation functions of the boundary theory. This is because we expect singularities whenever two points on the boundaries may be joined by a null geodesic \cite{Hubeny_2007,Gary:2009ae,maldacena2015lookingbulkpoint}. It would be interesting to relate the value of the stress tensor expectation value we computed to the locations of these singularities, as a prediction for the boundary theory.

    \item Our analysis could be extended to more general situations. First, it would be interesting to consider more general matter content that is not constrained by conformal symmetry. Second, one could consider boundaries at locations that are less symmetric. Finally, one could perform a similar calculation in higher dimensions, computing $\ev{T_{\mu\nu}}$ as \cite{Dowker:1975gt}
    \begin{align}
    \ev{T_{\mu\nu}(x)}=-i\lim_{x'\to x}\left[\nabla_\mu\nabla_{\nu'}-\frac{1}{2}g_{\mu\nu'}g^{\lambda\sigma'}\nabla_\lambda\nabla_{\sigma'}\right]G_F(x,x'),
    \end{align}
    where $G_F$ is the Feynman propagator, which may be calculated in the geodesic approximation. In $2+1$D, it would also be interesting to explore how a non-zero $\ev{T_{\mu\nu}}$ in the vacuum state affects boundary gravitons and valid embeddings of the boundary in de Sitter spacetime \cite{Marolf_2012}.
\end{enumerate}

\noindent{\bf Acknowledgements}
I am deeply grateful to Batoul Banihashemi, Dan Eniceicu, Mason Kamb, Albert Law, Henry Lin, G. Bruno de Luca, Aron Wall and Sungyeon Yang for helpful discussions. I am especially thankful to Gonzalo Torroba for many insightful comments, to Edgar Shaghoulian for suggesting the role of the Casimir effect, initial discussions, and helpful comments on the draft, and to Eva Silverstein for suggesting the exploration of the Gao-Wald theorem in this context, the role of ``moving mirror" radiation, and further guidance.

\appendix

\section{General solution to dilaton equations of motion}\label{sec:deoms}
In this section we outline the derivation of the general solution to the dilaton equations of motion given in Eq. (\ref{eq:dilaton_eom}):
\begin{align}
    &\left(-\partial_\varphi^2+\tan\sigma\partial_\sigma-\sec^2\sigma\right)\phi=2\pi T_{\sigma\sigma}, \label{eq:deq1}\\
    &\left(-\partial_\sigma^2+\tan\sigma\partial_\sigma+\sec^2\sigma\right)\phi=2\pi T_{\varphi\varphi},\label{eq:deq2}\\
    &\left(-\partial_\sigma\partial_\varphi+\tan\sigma\partial_\varphi\right)\phi=2\pi T_{\sigma\varphi}.\label{eq:deq3}
\end{align}
We have replaced $\ev{T_{\mu\nu}}$ by $T_{\mu\nu}$ everywhere.
The general solution to Eq. (\ref{eq:deq2}) is
\begin{align}
 \label{eq:dstep}   \phi=\phi_r\cos\varphi\sec\sigma+c_1(\varphi)\sec\sigma+c_2(\varphi)\tan\sigma+2\pi\sec\sigma\int^\sigma d\sigma' \sin\sigma'T_{\varphi\varphi}-2\pi\tan\sigma\int^\sigma d\sigma' T_{\varphi\varphi}.
\end{align}
Since we are looking for solutions symmetric under $\sigma\to -\sigma$, we can set $c_2=0$. Plugging the resulting general solution for $\phi$ above into Eq. (\ref{eq:deq1}) then gives
\begin{align}
    -(c_1''(\varphi)+c_1(\varphi))\sec\sigma+2\pi\tan\sigma\int^\sigma d\sigma' \partial_\varphi^2T_{\varphi\varphi}-2\pi\sec\sigma\int^\sigma d\sigma' \sin\sigma' (\partial_\varphi^2 T_{\varphi\varphi}+T_{\varphi\varphi})=2\pi T_{\sigma\sigma}.
\end{align}
The two integrals in the first line may be simplified using a combination of integration by parts and the equations
\begin{align}
    &\partial_\sigma(\cos\sigma T_{\sigma\sigma})=\cos\sigma\partial_\varphi T_{\varphi\sigma}-\sin\sigma T_{\varphi\varphi},\\
    &\partial_\varphi T_{\varphi\varphi}=\partial_\sigma T_{\sigma\varphi},
\end{align}
which are the two components of $\nabla_\mu T^{\mu\nu}=0$. These simplifications give the following differential equation for $c_1(\varphi)$:
\begin{align}
    c_1''(\varphi)+c_1(\varphi)=0.
\end{align}
The general solution to this equation which is symmetric under $\varphi\to -\varphi$ is $c_1(\varphi)=c_1\cos\varphi$. We can plug this into Eq. (\ref{eq:dstep}) to get the general solution for $\phi$:
\begin{align}
    \phi=\phi_r'\frac{\cos\varphi}{\cos\sigma}-2\pi\sec\sigma\int^\sigma d\sigma' \cos\sigma' \int^{\sigma'}d\sigma'' T_{\varphi\varphi}\nonumber,
\end{align}
where we defined $\phi_r'=\phi_r+c_1$.

\section{Computation of the acceleration of the boundaries}\label{sec:acceleration}
In this section we will show that the boundaries considered in Section \ref{sec:bdry2} are accelerating with an acceleration that increases with $\zeta$. Recall that the boundaries are given by the curves
\begin{align}
\label{eq:bdrypath}
    \frac{\cos\varphi}{\cos\sigma}=\tanh\zeta\implies \sin\rho_c=\tanh\zeta.
\end{align}
We may parametrize this boundary path as $\gamma^\mu=(t(\tau),\rho(\tau))$, which is fixed using Eq. (\ref{eq:bdrypath}) and the condition
\begin{align}
    -\cos^2\rho\left(\frac{dt}{d\tau}\right)^2+\left(\frac{d\rho}{d\tau}\right)^2=-1,
\end{align}
which says that the path is timelike. The four-acceleration is then given by
\begin{align}
    a^{\mu}=\frac{d^2\gamma^\mu}{d\tau^2}+\Gamma^\mu_{\rho\sigma}\frac{d\gamma^\rho}{d\tau}\frac{d\gamma^\sigma}{d\tau},
\end{align}
using which we can compute
\begin{align}
    a_\mu a^\mu=\sinh^2\zeta,
\end{align}
which increases on increasing $\zeta$.

\section{$\ev{T_{\mu\nu}}_g$ for boundaries at constant dilaton outside the static patches}\label{sec:tmunu}
In this section we outline in more detail the numerical results for $\ev{T_{\mu\nu}}_g$ in the case where boundaries are at locations of constant dilaton. The procedure for computing this quantity is outlined in Section \ref{sec:bdry2}, along with its functional form inside the two static patches. Here we collect figures displaying its functional form in global coordinates (\ref{eq:metric}) that cover the entire spacetime, and comment on some of their features. 

\begin{figure}
\centering
\begin{tabular}{c}
\subfloat[$T_{\sigma\sigma}(\sigma)$]{\includegraphics[width = 0.38\textwidth]{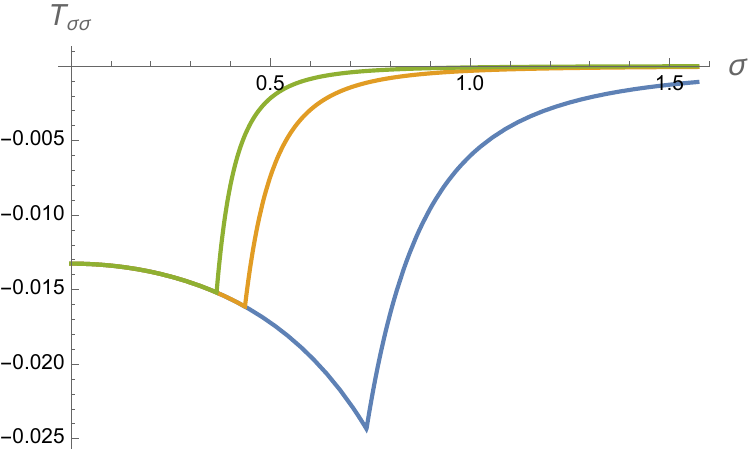}}
\subfloat[$T_{\varphi\varphi}(\sigma)$]{\includegraphics[width = 0.38\textwidth]{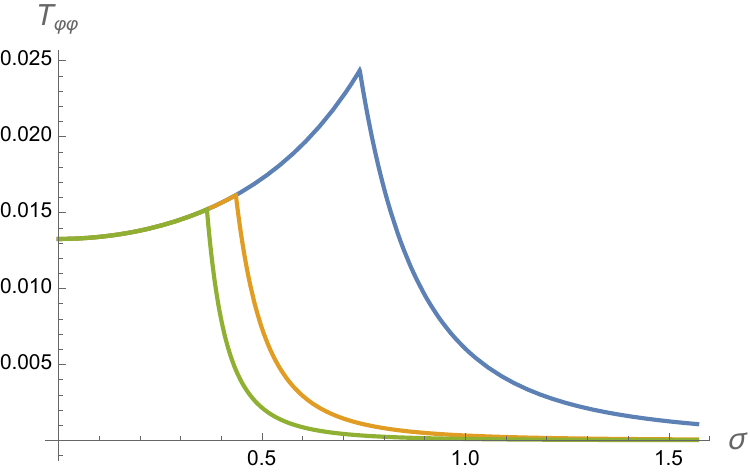}}\\
\subfloat[$T_{\sigma\sigma}(\varphi)$]{\includegraphics[width = 0.38\textwidth]{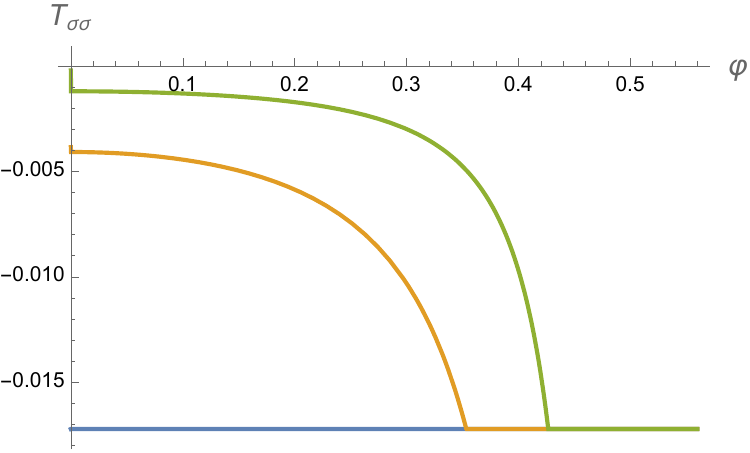}}
\subfloat[$T_{\varphi\varphi}(\varphi)$]{\includegraphics[width = 0.38\textwidth]{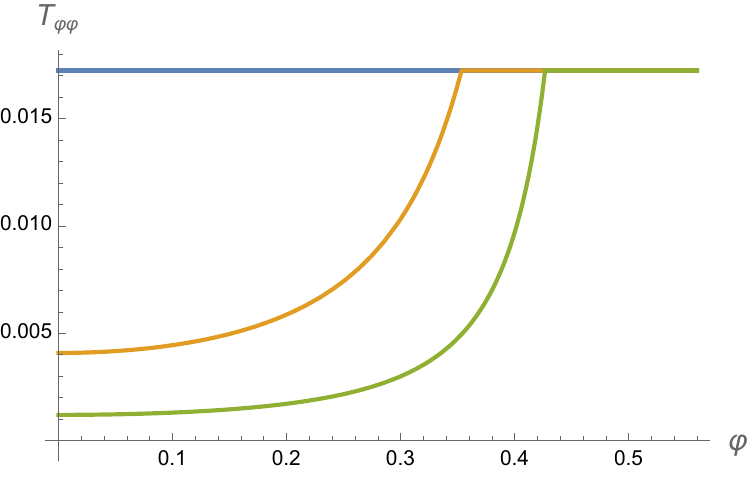}}
\end{tabular}
\caption{Plots of various components of $\ev{T_{\mu\nu}}$ in the $(\sigma,\varphi)$ coordinates, for $c=1$ and various values of $\zeta$. In blue is $\zeta=1$, in orange is $\zeta=1.7$, and in green is $\zeta=2$.}
\label{fig:Tsigmavarphi}
\end{figure}

In Figure \ref{fig:Tsigmavarphi} we plot $\ev{T_{\sigma\sigma}}$ and $\ev{T_{\varphi\varphi}}$ as a function of $\sigma$ and $\varphi$ for various values of $\zeta$. $\ev{T_{\sigma\varphi}}$ is approximately zero. We also see that $T_{\sigma\sigma}$ and $T_{\varphi\varphi}$ are negatives of each other. Notice that these quantities don't seem to depend on $\zeta$ inside the horizons, but depend on it outside the horizons starting at $\zeta$-dependent locations. In particular, as $\zeta$ increases, $T_{\sigma\sigma}$ becomes less negative. This is due to the increasing acceleration of the boundaries (as discussed in Appendix \ref{sec:acceleration}) and the resulting ``moving mirror" radiation. 

\begin{figure}
\centering
\begin{tabular}{c}
\subfloat[$T_{\sigma'\sigma'}(\sigma')$]{\includegraphics[width = 0.38\textwidth]{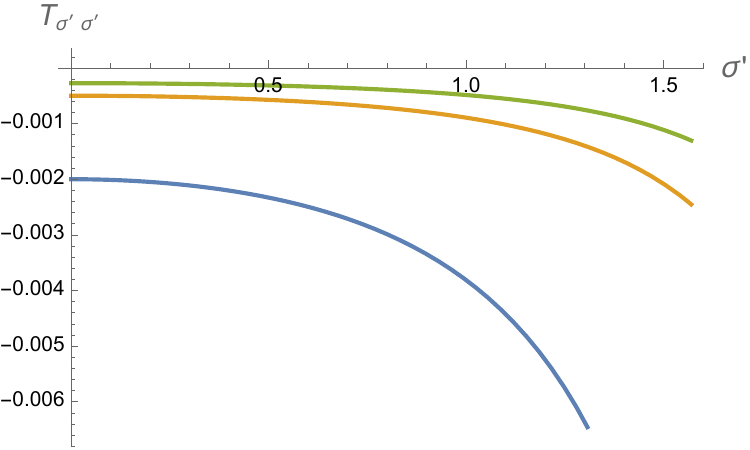}}
\subfloat[$T_{\varphi'\varphi'}(\sigma')$]{\includegraphics[width = 0.38\textwidth]{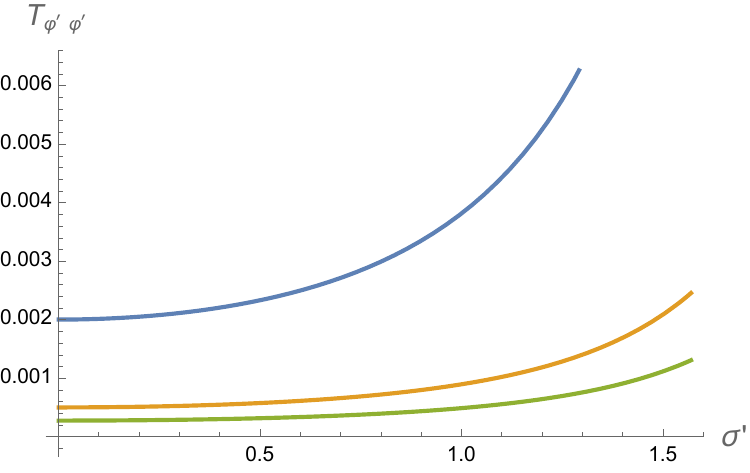}}\\
\subfloat[$T_{\sigma'\sigma'}(\varphi')$]{\includegraphics[width = 0.38\textwidth]{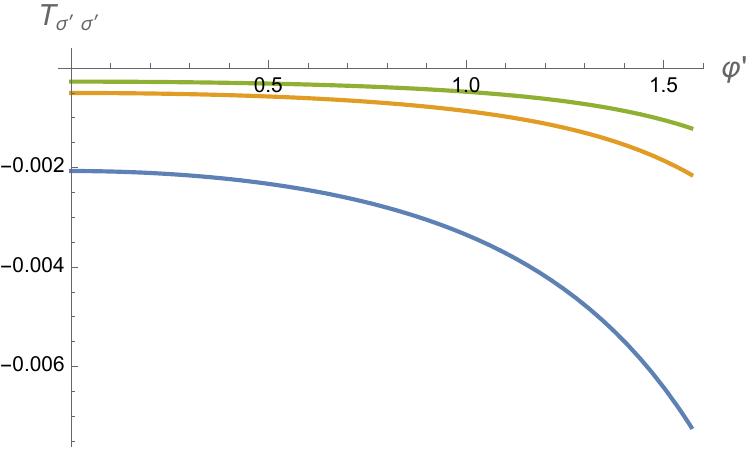}}
\subfloat[$T_{\varphi'\varphi'}(\varphi')$]{\includegraphics[width = 0.38\textwidth]{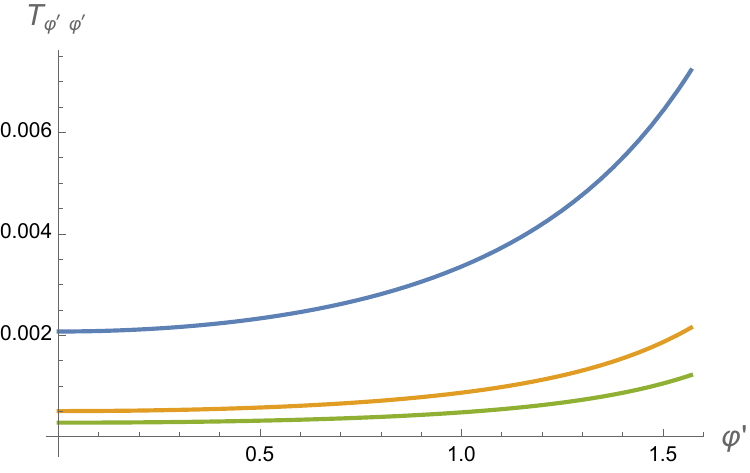}}
\end{tabular}
\caption{Plots of various components of $\ev{T_{\mu\nu}}$ in the $(\sigma',\varphi')$ coordinates, for $c=1$ and various values of $\zeta$. In blue is $\zeta=1$, in orange is $\zeta=1.7$, and in green is $\zeta=2$.}
\label{fig:Tsigmavarphi2}
\end{figure}

We also plot $\ev{T_{\sigma'\sigma'}}$ and $\ev{T_{\varphi'\varphi'}}$ as a function of $\sigma'$ and $\varphi'$ for various values of $\zeta$ in Figure \ref{fig:Tsigmavarphi2}. $\ev{T_{\sigma'\varphi'}}$ is approximately zero. Recall the $(\sigma',\varphi')$ coordinates are given by Eq. (\ref{eq:transformation}), with the boundaries at $\varphi_c'=\pm\frac{\pi}{2}$.

\bibliographystyle{JHEP}
\bibliography{main.bib}
\end{document}